\documentclass[twocolumn,prl]{revtex4}
\usepackage{graphicx}
\usepackage{amsmath}

\def\eq#1{Eq.~(\ref{#1})}
\def\fig#1{Fig.~\ref{#1}}

\newcommand{\boldv}[1]{{\mathbf #1}}

\newcommand{\ha}{\hspace{1cm}}

\newcommand{\rez}[1]{\frac{1}{#1}}
\newcommand{\grad}{{\mathbf \nabla}}

\newcommand{\mean}[1]{\langle #1 \rangle}

\begin{document}

\title{Efficiency of initiating cell adhesion in hydrodynamic flow}

\author{C. Korn}
\author{U. S. Schwarz}
\affiliation{University of Heidelberg, Im Neuenheimer Feld 293, D-69120 Heidelberg, Germany}
\affiliation{Max Planck Institute of Colloids and Interfaces, D-14424 Potsdam, Germany}

\begin{abstract}
  We theoretically investigate the efficiency of initial binding
  between a receptor-coated sphere and a ligand-coated wall in linear
  shear flow. The mean first passage time for binding decreases
  monotonically with increasing shear rate. Above a saturation
  threshold of the order of a few 100 receptor patches, the binding
  efficiency is enhanced only weakly by increasing their number and
  size, but strongly by increasing their height. This explains why
  white blood cells in the blood flow adhere through receptor patches
  localized to the tips of microvilli, and why malaria-infected red
  blood cells form elevated receptor patches (\textit{knobs}).
\end{abstract}

\maketitle

Cohesion in biological systems and biotechnological applications is
usually provided by specific bonds between receptors and ligands.  The
formation of these bonds requires a physical transport process which
brings receptors and ligands to sufficient proximity for binding. On
the cellular level, one of the most prominent examples is the binding
of blood-born cells to the vessel walls under the conditions of
hydrodynamic flow. For white blood cells, initial binding to the
vessel walls is the first step in their hunt for pathogens, which is
then followed by rolling adhesion, firm arrest and extravasation
\cite{c:spri94}. Similar processes are used by stem and cancer cells
which travel the body with the blood stream.  Initiating binding to
vessel walls is also essential for malaria-infected red blood cells in
order to avoid clearance by the spleen and possibly also to foster
rupture and release of new parasites into the blood stream
\cite{c:naga00}. Similar questions about initial binding under flow
conditions arise for bacteria, e.g.\ when binding to the intestinal
wall \cite{c:thom02}, and in biotechnological applications, e.g.\ for
adhesion-based cell sorting \cite{c:fore04}. In order to control shear
flow and cell density in a quantitative way, a standard setup are flow
chambers \cite{c:alon95}. An essential but largely unexplored aspect
of these processes is receptor geometry, that is size, height and
separation distance of the receptor patches. One way to address this
issue experimentally is the use of receptor-coated beads
\cite{c:pier02}.

Earlier theoretical efforts in this context have been focused mainly
on issues related to white blood cells, including modelling of the
initiation of adhesion at high cell densities (e.g.\ due to
hydrodynamic interactions) \cite{c:king01} and the process of
rolling adhesion \cite{c:chan00}. In these studies, the parameters
characterizing receptor-ligand binding are usually fixed at
physiologically motivated values. In this Letter, we take a more
general view and ask how a variable receptor geometry affects cell
capture in hydrodynamic flow. In order to study this issue in a
systematic way, we spatially resolve receptors and ligands.
The cell is modeled as a rigid spherical Brownian particle
in linear shear flow carrying receptors for ligands covering a planar
boundary wall. In the absence of interactions with other particles or
external forces, there is no reason for such a particle to drift
towards the wall and initial binding has to rely completely on thermal
diffusion. In order to arrive at a generic model, we consider the
simplest of the possible downward driving forces in experiments with
flow chambers, namely gravity. The particle is set free at a certain
height above the wall and we calculate the mean first passage time
(MFPT) for the first receptor-ligand encounter as a measure for the
efficieny of initial cell binding. We consider three models of
increasing complexity in regard to the spatial distribution of
receptors and ligands. We first show that if the receptors on the cell
and the ligands on the substrate are distributed homogeneously, then
the corresponding MFPTs can be calculated exactly.  In the case that
receptors are spatially resolved, we use extensive computer
simulations to calculate the MFPTs as a function of their number and
spatial dimensions, both in two and three dimensions.  As a third
case, we in addition consider spatially resolved ligand distribution.

\begin{figure}
\begin{center}
\includegraphics[width=\columnwidth]{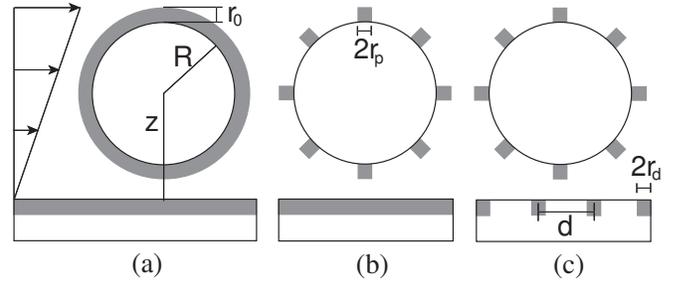}
\caption{A cell of radius $R$ moves in linear shear flow above a wall
  at a height $z$, subject to hydrodynamic, gravitational and
  thermal forces.  The efficiency for initiating binding is assessed
  by calculating the mean first passage time for approach of receptor
  and ligand to a capture distance $r_0$, that is for overlap of the
  gray regions.  Three models of increasing complexity are studied
  here: (a) homogeneous coverage of cell and wall, (b) $N_r$
  equidistantly placed receptor patches on the cell, each with radius
  $r_p$, and (c) ligand patches of radius $r_d$ separated by
  distance $d$.}
\label{cartoon}
\end{center}
\end{figure}

\fig{cartoon} introduces the parameters of our model. We consider a
sphere of radius $R$ which moves with the hydrodynamic flow in
positive x-direction at a height $z$ above the wall with normal
$\boldv{e}_z$. The simplest possible flow pattern is linear shear flow
with shear rate $\dot \gamma$.  With the no-slip boundary condition at
the wall, the unperturbed velocity profile reads $\boldv{u}^\infty =
\dot \gamma z \boldv{e}_x$.  For a typical cell radius $R = 5\ \mu$m
and a typical shear rate $\dot \gamma =$ 100 Hz, the Reynolds number
in aqueous solution is well below $1$ and the hydrodynamic flow is
essentially described by the Stokes equation for incompressible
fluids. This is even more true for smaller particles like micron-sized
beads. Scaling estimates also show that for typical parameter values
for cell elasticity, deformations due to shear flow and lubrication
forces are small and therefore the spherical approximation is
justified. In addition to the hydrodynamic forces, in our model there
are also gravitational and thermal forces acting on the particle. Such
a combination of forces is the subject of Stokesian Dynamics
\cite{c:erma78} and in our case leads to the following
Langevin equation \cite{uss:korn06b}:
\begin{align}
  \label{langevin-01}
  \partial_t \boldv{X}_t = \boldv{u}^\infty + 
  \mathsf{M} (\boldv{F}^G + \boldv{F}^{S})
  + k_B T_a \mathsf{B}\grad\mathsf{B}^T + \boldv{g}^S_t\ .
\end{align}
$T_a$ is ambient temperature. The gravitational force reads
$\boldv{F}^G = - g \Delta \rho (4 \pi R^3/3) \boldv{e}_z$ with some
density difference $\Delta \rho$. The subscript $t$ denotes random
variables. As usual, the thermal force $\boldv{g}_t$ is assumed to be
Gaussian:
\begin{align}
  \label{noise}
  \mean{\boldv{g}_t} = 0\ , \ha \mean{\boldv{g}_t \boldv{g}_{t'}} =
  2 k_B T_a \mathsf{M}\delta(t - t')\ .
\end{align}
The superscript $S$ for the multiplicative noise term in
\eq{langevin-01} indicates that it has to be interpreted in the usual
Stratonovich sense.  The matrix $\mathsf{B}$ in \eq{langevin-01} is
related to the mobility matrix $\mathsf{M}$ through $\mathsf{M} =
\mathsf{BB}^T$. Vectors in \eq{langevin-01} are six-dimensional,
representing the spatial and orientational degrees of freedom. The
mobility matrix $\mathsf{M}$ and the shear force $\boldv{F}^{S}$ for a
spherical particle above a wall cannot be obtained in analytically
closed form. However, they can be calculated numerically to high
accuracy and we will use this for our simulations
\cite{c:perk92}.

Considering the physical dimensions of our problem shows that the
motion of the cell is essentially governed by two dimensionless
numbers.  For length, the natural scale is cell radius $R$. Mobility
and shear force scale as $M = 1/(6 \pi \eta R)$ and $F^S = 6 \pi \eta
R^2 \dot \gamma$, respectively. For time, there are two relevant time
scales, the deterministic time scale $1/\dot \gamma$ and the diffusive
time scale $R^2 / D = 6 \pi \eta R^3 / k_B T_a$, where we have used
the Einstein relation $D = M k_B T_a$ for the diffusion constant $D$.
Therefore the relative importance of hydrodynamic to thermal motion is
described by the \textit{P\'eclet number} $Pe = 6 \pi \eta R^3 \dot
\gamma / k_B T_a$.  In the limits $Pe \to 0$ and $Pe \to \infty$,
diffusive and deterministic motion dominate, respectively. The
gravitational force introduces another dimensionless number, which we
call the \textit{P\'eclet number in z-direction}, $Pe_z = F^G R / k_B
T_a = 4 \pi g \Delta \rho R^4 / (3 k_B T_a)$.  In the following, we
will non-dimensionalize length and time by $R$ and $6 \pi \eta R^3 /
k_B T_a$, respectively.

We start by considering homogeneous coverage of cell and wall with
receptors and ligands, respectively, compare \fig{cartoon}a. Then
rotational degrees of freedom are irrelevant. Because the wall
breaks the symmetry only in the z-direction, motion in the x-y-plane
is decoupled from our problem.  Thus in this case we essentially deal
with a MFPT in one dimension, which is independent of shear rate $\dot
\gamma$ and which can be approached with standard methods for the
appropriate Fokker-Planck equation. Binding is identified with
approach of receptor and ligand to a capture distance $r_0$. Applied
to the case of homogeneous coverage, the cell has to fall to the
capture height $1 + r_0$. If dropped from the initial height $z_0$,
the respective MFPT can be shown to be
\begin{align}
  \label{z-mfpt}
  T_h = \rez{Pe_z} \int_{1+r_0}^{z_0} dz \frac{1}{M_{zz}(z)}\ .
\end{align}
Thus the MFPT scales inversely with the gravitational force driving
the cell onto the wall. With the lubrication approximation $M_{zz}(z)
\approx 1 - 1/z$ we find
\begin{align}
  \label{z-mfpt-lub}
  T_h \approx \rez{Pe_z} \left[z_0 - 1 - r_0 + \ln\left(\frac{z_0 - 1}{r_0}\right)\right]\ .
\end{align}
Thus the MFPT diverges logarithmically with vanishing capture distance
$r_0$ (that is when the cell has to get infinitely close to the wall)
and linearly with initial height $z_0$ (that is when the cell starts
infinitely far away from the wall). Although only the constant
force chosen here results in an analytical result like \eq{z-mfpt-lub},
for other types of force laws it is straightforward to numerically
calculate corresponding falling times $T_h$ \cite{uss:korn06b}.

\begin{figure}
\begin{center}
\includegraphics[width=\columnwidth]{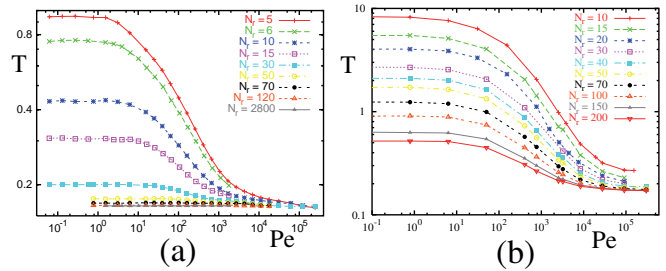}
\caption{Effect of shear rate on mean first passage time $T$ for initial
  binding for spatially resolved receptors in (a) two (2D) and (b) three
  dimensions (3D). The scales for length and time are $R$ and $6 \pi \eta
  R^3 / k_B T_a$, respectively. The P\'eclet number $Pe = 6 \pi \eta R^3
  \dot \gamma / k_B T_a$ and the P\'eclet number in z-direction $Pe_z =
  4 \pi g \Delta \rho R^4 / (3 k_B T_a)$ represent the strengths of the
  hydrodynamic and gravitational forces, respectively, relative to the
  thermal forces. $z_0 = 2$, $Pe_z = 50$, $r_0 = r_p = 10^{-3}$.}
\label{peclet}
\end{center}
\end{figure}

We next consider the case of a spatially resolved receptor
distribution, compare \fig{cartoon}b.  Now the cell is equidistantly
covered with $N_r$ receptor patches, each with radius $r_p$ and height
$r_0$. We first note that in this case, initial orientation becomes
important. Moreover now shear rate $\dot \gamma$ enters the analysis:
the shear flow increases cell rotation, and for
heterogeneous receptor coverage, this strongly influences when the
first receptor can bind the first ligand. Because experimentally it is
hardly possible to prepare the initial orientation of the cell, in the
following we average over all possible initial orientations.  One can
show that the angle-averaged MFPT is the MFPT to fall from initial
height $z_0$ to some intermediate height $z_m$ according to
\eq{z-mfpt} (that is independent of orientation) plus the
angle-averaged MFPT to bind from the initial height $z_m$. In this
sense, the initial height is not relevant for our problem and in the
following we always use $z_0 = 2$, that is the cell has to fall for
the distance of one radius before binding can occur.

In \fig{peclet}a and b we show the MFPT as obtained by extensive
computer simulations as a function of P\'eclet number $Pe \sim \dot
\gamma$ and receptor patch number $N_r$ for two (2D) and three
dimensions (3D), respectively. Here 2D mean that translational motion
is restricted to the x-z-plane and rotations are restricted about the
y-axis, which allows for much faster simulations. Each data point in
\fig{peclet} is the average of at least $10^5$ simulated trajectories
of the Langevin equation \eq{langevin-01}. Our simulations are very
time-consuming because with the receptor patches we resolve objects of
typical size $10^{-3}$, that is nm-sized patches on micron-sized
cells. From \fig{peclet} we first note that $T$ decreases
monotoneously with increasing $Pe$ and that the shear rate does not
change the relative sequence of the curves for different $N_r$. Thus
the larger shear rate and the more receptor patches present, the more
efficient cell capture. The crossover between the diffusion- and
convection-dominated regimes does not occur at $Pe \approx 1$, but at
much larger values $Pe \approx 10^2$. Next we note that in the 2D case
(\fig{peclet}a), for large $Pe$ or large $N_r$, all curves level off
to the exact result for homogeneous coverage from \eq{z-mfpt}, because
in these two limits, the binding process effectively becomes
rotationally invariant. In the 3D-case (\fig{peclet}b), the
homogeneous reference value is only achieved for large receptor
numbers. The reason is that for small numbers of receptors, the cell
might have to rotate around the x-axis before a receptor moving on a
circle parallel to the x-z-plane is able to bind a ligand on the wall.
Therefore in 3D thermal diffusion remains essential even in the case
of large P\'eclet number.

\begin{figure}
\begin{center}
\includegraphics[width=\columnwidth]{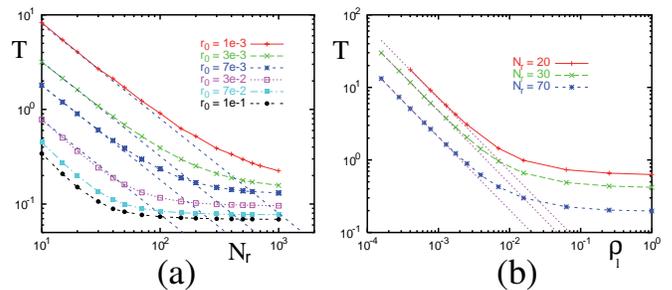}
\caption{Mean first passage time $T$ in 3D in the diffusive limit ($Pe
  \approx 0$) as a function of (a) the number of receptor patches
  $N_r$ for different values of patch height $r_0$ and (b) ligand
  density $\rho_l$ as varied by decreasing ligand patch distance $d$
  for different values of the number of receptors $N_r$. The dashed
  lines show the scaling behaviours $T \sim 1/N_r$ and $T \sim
  1/\rho_l$ at low receptor and ligand coverage, respectively.  $z_0 =
  2$, $Pe_z = 50$, $r_0 = r_p = r_d = 10^{-2}$.}
\label{geometry}
\end{center}
\end{figure}

In order to achieve a better understanding of the simulation results
shown in \fig{peclet}, it is instructive to decompose the process into
periods of falling and rotation, respectively. A detailed analysis
shows that in the 2D-case this decomposition allows to derive
scaling laws for different limits in regard to $Pe$ and $Pe_z$. An
important case is the one of large $Pe_z$, when the cell is strongly
driven onto the wall. Then the binding process can be decomposed into
a initial falling period described by \eq{z-mfpt}, followed by a
purely rotational search for ligand, which is independent of $Pe_z$
and can be calculated analytically:
\begin{equation}
\label{th-mfpt}
  T_r = \frac{A_\theta \Delta \theta^2 \coth(\frac{A_\theta \Delta \theta}{2D_\theta}) 
    - 2D_\theta \Delta \theta}{2 A_\theta^2 \theta_s}\ .
\end{equation}
Here $\Delta \theta \approx \theta_s$ is the angle between the
absorbing boundaries, $\theta_s = 2 \pi / N_r$ the angle between
receptor patches, $A_\theta \approx Pe / 2$ the rotational drift and
$D_\theta \approx 3/4$ the rotational diffusion constant. From
\eq{th-mfpt} we get $T \sim 1/N_r^2$ and $T \sim 1/(N_r Pe)$ for small
and large $Pe$, respectively, in excellent agreement with the scaling
found in our simulations. In general, \eq{th-mfpt} is a good
qualitative description of the 2D data shown in \fig{peclet}.

In order to consider the case of spatially resolved ligand, compare
\fig{cartoon}c, we cover the boundary wall with a square lattice of
circular ligand patches, with lattice constant $d$ and patch radius
$r_d$. Because again the P\'eclet number does not change the relative
sequence of the different MFPT curves, the efficiency of initial cell
binding as a function of receptor and ligand geometry can be
investigated in the diffusive limit $Pe \approx 0$. \fig{geometry}
shows for this case that the MFPT saturates both when increasing
receptor coverage by increasing $N_r$ or ligand coverage by decreasing
$d$. A similar saturation behaviour is also found when increasing
receptor and ligand patch sizes $r_p$ and $r_d$.  Typically the
saturation threshold for the parameters used is located at a mean
patch-to-patch distance $d$ of about $0.17$, both in regard with
receptors and ligands. For a coverage below the threshold,
\fig{geometry} shows that the MFPT scales like $\sim 1/N_r \sim d^2$
and $\sim 1/\rho_l \sim d^2$ in regard to receptor and ligand
coverage, respectively, where $d$ represents the distance between
receptor and ligand patches, respectively.  This can be understood by
noting that the 1D MFPT for capture by diffusion scales $\sim d^2$
where $d$ is the distance between the two absorbing boundaries. The
saturation effect observed with respect to $N_r$, $\rho_l$, $r_p$ and
$r_d$ results from the space-filling nature of diffusion which has
been implicated before for the efficiency of ligand capture by a cell
\cite{c:berg77}.

\begin{figure}
\begin{center}
\includegraphics[width=0.7\columnwidth]{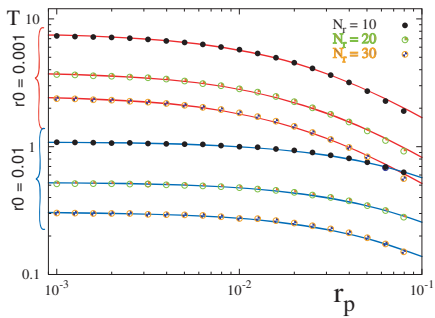}
\caption{Mean first passage time $T$ in 3D in the diffusive limit ($Pe
  \approx 0$) as a function of the receptor patch radius $r_p$ for two
  different values of $r_0$ and three different values of $N_r$.  For
  better comparison with regard to $r_0$ the mean first passage time
  $T_h$ for homogeneous coverage from \eq{z-mfpt} has been
  substracted. The lines are fits according to \eq{receptor-fit}.
  Other parameters as in \fig{peclet}.}
\label{height}
\end{center}
\end{figure}

We finally turn to the effect of the capture distance $r_0$.  In
\fig{height} we show the MFPT in the diffusive limit as a function of
the receptor patch radius $r_p$ for $r_0 = 10^{-3}$ and $r_0 =
10^{-2}$ as well as for three different values of $N_r$. We first note
that the MFPT is much more influenced by a change in $r_0$ or $N_r$
then by a change in $r_p$. One can show on geometrical grounds that
the two parameters $r_0$ and $r_p$ conspire to define an effective
receptor patch size $\sqrt{r_0} + r_p$ which then in turn determines
the probability for binding. This line of reasoning leads to the
formula
\begin{align}
  \label{receptor-fit}
  T = \frac{a}{b + r_p} + T_h
\end{align}
where $a = 2 t_d/\sqrt{r_0}$, $b = \sqrt{r_0} / 2$, $t_d$ is a typical
diffusion time between binding attempts which scales $\sim 1/N_r$, and
$T_h$ is the homogeneous result from \eq{z-mfpt}. \eq{receptor-fit}
implies that even for vanishing receptor size the MFPT remains finite
due to a finite $r_0$. The lines in \fig{height} show that this
equation can be fitted extremely well to the simulation results.
Moreover the fitted values for $a$ and $b$ agree roughly with the
predicted scaling behaviour.

In summary, our results show that the efficiency for initiating cell
adhesion in hydrodynamic flow is strongly enhanced by increasing the
number of receptor patches $N_r$, but only up to a saturation
threshold.  An increase of patch size $r_p$ leads only to a weak
enhancement of binding efficiency.  In contrast, a strong enhancement
results from increasing the patch height $r_0$.  For example, for a
few hundred receptor patches, an elevation of $r_0 = 10^{-2}$
makes initial binding already as efficient as for a homogeneously
covered cell. Strikingly, white blood cells are indeed characterized
by such a receptor geometry, because they are covered with hundreds of
protrusions (\textit{microvilli}, typical height 300 nm, corresponding
to $r_0 = 0.06$) which carry adhesion receptors like L-selectin at
their narrow tips \cite{c:alon95}. In general, white blood cells
operate in the limit of a homogeneously covered cell not only due to
their receptor geometry, but also because during capture they are
usually exposed to environments with $Pe \approx 10^4 - 10^5$ (a
typical value for $Pe_z$ is $300$).  The principle of enhancing capture
efficiency by elevation of receptor patches seems to be also used by
other biological systems.  One example of large medical relevance
appears to be malaria-infected red blood cells, which develop
thousands of little adhesive protrusions (\textit{knobs}, typical
height 20 nm, corresponding to $r_0 = 0.004$)
\cite{c:naga00}. The results presented here do not only allow
to understand the efficiency of cell capture in these biological
systems in a unified way, but can also be used for developing
corresponding applications in biotechnology, including adhesion-based
cell or particle sorting.

This work was supported by the Emmy Noether Program of the German
Research Foundation (DFG) and the Center for Modelling and Simulation
(BIOMS) at Heidelberg.


\begin{thebibliography}{19}
\bibitem[{\citenamefont{Korn and Schwarz}(2006)}]{uss:korn06b}
A detailed derivation of the Langevin equation and more
details of our simulation algorithms and results will be published elsewhere.
\end{thebibliography}

\end{document}